\documentclass[dvipdfmx]{article}
\usepackage{latexsym}
\usepackage{amsmath,amssymb}
\usepackage[hiresbb]{graphicx}
\usepackage{tikz}
\usepackage{fancybox,ascmac}
\usepackage{bm}
\usepackage{mathtools,slashed}
\usepackage{fancyhdr}

\begin{document}

\begin{flushright}
DPUR/TH/78\\
January, 2024\\
\end{flushright}
\vspace{20pt}

\pagestyle{empty}
\baselineskip15pt

\begin{center}
{\Large\bf  Effective Potential for Conformal Factor and GL(4) Symmetry \vskip 1mm }

\vspace{20mm}

{\large Ichiro Oda\footnote{
           E-mail address:\ ioda@sci.u-ryukyu.ac.jp
                  }

\vspace{10mm}
           Department of Physics, Faculty of Science, University of the 
           Ryukyus,\\
           Nishihara, Okinawa 903-0213, Japan\\          
}

\end{center}


\vspace{10mm}
\begin{abstract}

We revisit the issue that the effective potential for the conformal factor of the metric, which is generated by
quantized matter fields, possesses a non-vanishing vacuum expectation value (VEV) or not.  We prove that the effective
potential has a vanishing vacuum expectation value on the basis of a global $GL(4)$ symmetry. We also account for the
reason why there seem to be two different effective potentials for the conformal factor in a theory, one of which
gives rise to a vanishing VEV for the conformal factor whereas the other does a non-vanishing VEV.  

\end{abstract}

\newpage
\pagestyle{plain}
\pagenumbering{arabic}


\section{Introduction}

There have been thus far a lot of studies of conformal factor or conformal mode of the metric tensor field. The conformal
factor is in essence a ghost mode in quantum field theory, which leads to negative energy at classical level and violates 
the unitarity at quantum level.  Thus, understanding the conformal factor might give us some clues for problems 
associated with ghost fields such as the massive ghost in higher-derivative gravities. 

The study of the conformal factor has been done mainly from three physical reasons:
The first reason is that in the Euclidean approach of general relativity, the Einstein-Hilbert action is unbounded from below
because of the conformal factor. The pragmatic approach for this issue is that the integration over the conformal 
factor is rotated in the complex plane so as to make the integral be formally convergent \cite{Cambridge}. 

The second reason comes from the well-known cosmological constant problem \cite{Weinberg}. For instance, starting with a theory with 
a metric $g_{\mu\nu}$ and a scalar field $\phi$, one can always construct a new gravitational theory with a new metric 
$\tilde g_{\mu\nu}$ and a new scalar field $\tilde \phi$ by performing a Weyl transformation $\tilde g_{\mu\nu} 
= \rho^2 g_{\mu\nu}$ and $\tilde \phi = \rho^{-1} \phi$ where $\rho$ is nothing but the conformal factor. It can be 
conjectured that some dynamical mechanism might give us an appropriate vacuum expectation value for the conformal 
factor in such a way that the cosmological constant is adjusted to be almost zero \cite{Peccei, Barr, Coughlan, Buchmuller, 
Tomboulis, Antoniadis}.     

The final reason is related to the problem of the degenerate metric \cite{Percacci1, Percacci2, Oda-Kin1, Oda-Kin2}. 
This problem is one of motivations behind the present work, so let us explain it in detail. In classical gravitational theories, 
the metric, or equivalently the vierbein, is usually postulated to be non-degenerate in order to guarantee the existence 
of its inverse metric or inverse vierbein. On the other hand, in quantum gravity, the degenerate metric is expected to arise 
since the change of the space-time topology occurs through such the degenerate metric \cite{Tseytlin, Horowitz}.  

It might be then useful to calculate the effective potential for the conformal factor of the metric since the nonvanishing 
vacuum expectation value of the conformal factor means that we have a nondegenerate metric at low energies even if 
there is a degenerate metric at high energies. Actually, this calculation has been performed so far 
in Refs. \cite{Percacci1, Percacci2, Oda-Kin1, Oda-Kin2}, 
but there seem to exist some confusions, in particular, when we start with a classical matter action in a curved background 
and calculate the one-loop effective potential for the conformal factor from quantized matter fields, we arrive at two different 
effective potentials. The origin of the two different effective potentials lies in the fact that we have two different background 
metrics in the cutoff regularization. In this article, we would like to clarify which effective potential is correct if we would
like to understand the problem of the degenerate metric.  

The outline of this article is as follows: In the next section we review calculations of the effective potential for the conformal 
factor of the metric from a quantized scalar field. In Section 3, we carry out similar calculations in case of a a quantized spinor field.
In Section 4, on the basis of a global $GL(4)$ symmetry we prove that the effective potential for the conformal factor must take the form 
of $V_{\rm{eff}} (g, \varphi) = \sqrt{-g} V (\varphi)$. The final section is devoted to a conclusion.

\section{Bosonic scalar fields}

We begin by reviewing calculations of the effective potential for the conformal factor of the metric from both quantized scalar
and spinor fields.
First, let us start with an action of a massive real scalar field in a curved background\footnote{We follow the notation and
conventions for the signature of the metric and the definition of the Riemannian tensors adopted in the MTW 
textbook \cite{MTW}.}
\begin{eqnarray}
S_B (\varphi, g_{\mu\nu} ) &=& - \frac{1}{2} \int d^4 x \sqrt{-g} 
( g^{\mu\nu} \partial_\mu \varphi \partial_\nu \varphi + m^2 \varphi^2 )
\nonumber\\
&=& - \frac{1}{2} \int d^4 x \sqrt{-g} 
\varphi ( - \Delta_g + m^2 ) \varphi,
\label{Scalar action}  
\end{eqnarray}
where we have defined the d'Alembertian operator $\Delta_g$ as
\begin{eqnarray}
\Delta_g = \frac{1}{\sqrt{-g}} \partial_\mu ( \sqrt{-g} g^{\mu\nu} \partial_\nu ).
\label{Delta}  
\end{eqnarray}
It is well-known that from this classical action one can derive the following effective action:
\begin{eqnarray}
\Gamma (g_{\mu\nu} ) &=& \frac{i}{2} {\rm{tr}} \log ( - \Delta_g + m^2 )
\nonumber\\
&=& \int d^4 x \sqrt{-g} \Bigl[ - \Lambda + \frac{M_{Pl}^2}{2} R + {\cal{O}} (R^2) \Bigr],
\label{Effective action}  
\end{eqnarray}
with the Planck mass squared being defined by $M_{Pl}^2 = \frac{1}{8 \pi G_N}$ where 
$G_N$ is the Newton's constant. Then, with $g_{\mu\nu} = \rho^2 \eta_{\mu\nu}$ this effective action 
gives rise to the effective potential for the constant conformal factor $\rho$:
\begin{eqnarray}
V_{\rm{eff}} = \Lambda \rho^4,
\label{Effective potential}  
\end{eqnarray}
where the effective potential $V_{\rm{eff}}$ is defined via $\Gamma (g_{\mu\nu} ) = -  \int d^4 x V_{\rm{eff}}$.
This is the stardard and well-known result, which was obtained in the induced gravity \cite{Akama}.

On the other hand, there is a different calculation of the effective potential for the conformal factor,
which will be described in what follows.  Provided that we define a conformal factor $\rho(x)$ by 
\begin{eqnarray}
g_{\mu\nu} = \rho(x)^2 \bar g_{\mu\nu},   \qquad
\varphi = \rho(x)^{-1} \bar \varphi,
\label{Conformal factor}  
\end{eqnarray}
where $\bar g_{\mu\nu}$ and $\bar \varphi$ are respectively a fixed fiducial metric and a fixed background,
the action (\ref{Scalar action}) can be written as
\begin{eqnarray}
S^\prime_B (\bar \varphi, \bar g_{\mu\nu}, \rho ) = - \frac{1}{2} \int d^4 x \sqrt{- \bar g} 
( \bar g^{\mu\nu} \partial_\mu \bar \varphi \partial_\nu \bar \varphi + m^2 \rho^2 \bar \varphi^2 
+ \dots),
\label{Scalar action 2}  
\end{eqnarray}
where the ellipses denote terms including derivatives of $\rho$. These terms are irrelevant
to our later argument  since we consider only constant $\rho$.  The latter definition in (\ref{Conformal factor}),
i.e., $\varphi = \rho(x)^{-1} \bar \varphi$ is needed to make the kinetic term for the scalar field be in the canonical form. 
Also note that the two actions (\ref{Scalar action}) and (\ref{Scalar action 2}) are equivalent at least classically.

Based on the action (\ref{Scalar action 2}), let us evaluate an one-loop effective potential for 
a constant conformal factor $\rho$ in a flat Minkowski background $\bar g_{\mu\nu} = \eta_{\mu\nu}$.
The partition function is defined as\footnote{In this calculation, we have a Jacobian factor
of form $\rho \delta^4 (0)$ which is set to be zero by using dimensional regularization procedure.}
\begin{eqnarray}
Z ( \rho ) &=& \int {\cal{D}} \bar \varphi \, {\rm{e}}^{i S^\prime_B}
= [ \det ( - \Box + m^2 \rho^2 ) ]^{-\frac{1}{2}}
\nonumber\\
&=& {\rm{e}}^{ -\frac{1}{2} {\rm{tr}} \log ( - \Box + m^2 \rho^2 ) },
\label{Partition}  
\end{eqnarray}
where $\Box = \eta^{\mu\nu} \partial_\mu \partial_\nu$. Then, an one-loop effective action is defined as
\begin{eqnarray}
\Gamma_B ( \rho ) &=& \frac{i}{2} {\rm{tr}} \log ( - \Box + m^2 \rho^2 )
\nonumber\\
&=& \frac{i}{2} \int d^4 x \langle x | \log ( - \Box + m^2 \rho^2 ) | x \rangle 
\nonumber\\
&=& \frac{i}{2} \int d^4 x \int \frac{d^4 p}{(2 \pi)^4} \log ( p^2 + m^2 \rho^2 )
\nonumber\\
&\equiv& - \int d^4 x \, V_{\rm{eff}}^{(B)},
\label{Effective action B}  
\end{eqnarray}
where $V_{\rm{eff}}^{(B)}$ indicates the one-loop effective potential for the conformal factor $\rho$.

The effective potential can be computed as follows:
\begin{eqnarray}
V_{\rm{eff}}^{(B)} &=& - \frac{i}{2} \int \frac{d^4 p}{(2 \pi)^4} \log ( p^2 + m^2 \rho^2 )
\nonumber\\
&=& \frac{1}{2} \int \frac{d^4 p_E}{(2 \pi)^4} \log ( p_E^2 + m^2 \rho^2 )
\nonumber\\
&=& \frac{1}{32 \pi^2} \int_0^{\Lambda^2} d p_E^2 \, p_E^2 \log ( p_E^2 + m^2 \rho^2 ), 
\label{Effective potential B}  
\end{eqnarray}
where we have performed the Wick rotation $p_0 = i p_4$ and defined $p_E^2 = p_1^2 + p_2^2
+ p_3^2 + p_4^2$.  By using the mathematical formula
\begin{eqnarray}
\int_0^{\Lambda^2} d x \, x \log ( x + c ) 
= \frac{1}{2} ( \Lambda^4 - c^2 ) \log ( \Lambda^2 + c ) - \frac{1}{4} \Lambda^4
+ \frac{1}{2} c \Lambda^2 + \frac{1}{2} c^2 \log c,
\label{Math}  
\end{eqnarray}
where $c$ being a constant, the one-loop effective potential for the conformal factor takes the form
\begin{eqnarray}
V_{\rm{eff}}^{(B)} = \frac{1}{32 \pi^2} \Bigl[ \frac{1}{2} m^4 \rho^4 \Bigl( \log \frac{m^2 \rho^2}{\Lambda^2} 
- \frac{1}{2}  \Bigr) + m^2 \Lambda^2 \rho^2 \Bigr], 
\label{Effective potential B2}  
\end{eqnarray}
where constant terms and terms including ${\cal{O}} (\frac{1}{\Lambda^2} )$ are subtracted. Adding suitable counterterms 
of the form $\Lambda^2 \rho^2$ and $\rho^4 \log \Lambda$ one can obtain the renormalized one-loop effective 
potential for the conformal factor
\begin{eqnarray}
V_{\rm{eff}}^{(B)} = \frac{1}{64 \pi^2} m^4 \rho^4 \Bigl( \log \frac{m^2 \rho^2}{\mu^2} - \frac{1}{2}  \Bigr), 
\label{Ren-effective potential}  
\end{eqnarray}
where $\mu$ is a renormalization mass \cite{Percacci2}. From this effective potential, we can see that the minimum of 
the renormalized one-loop effective potential (\ref{Ren-effective potential}) occurs for a nonzero $\rho$, thereby implying 
that the metric or the vierbein is nondegenerate and the inverse metric $g^{\mu\nu}$ or vierbein $e_a^\mu$ exists.
  
Here an important remark is in order. In deriving the effective potential (\ref{Ren-effective potential}), 
by means of the metric $\bar g^{\mu\nu}$, we have adopted a momentum cutoff, $\bar g^{\mu\nu} p_{E\mu} p_{E\nu} 
= \delta^{\mu\nu} p_{E\mu} p_{E\nu} < \Lambda^2$. On the other hand, if we use a different momentum cutoff 
by using the metric $g^{\mu\nu}$, i.e., $g^{\mu\nu} p_{E\mu} p_{E\nu} = \rho^{-2} \delta^{\mu\nu} p_{E\mu} p_{E\nu} 
< \Lambda^2$, a similar calculation provides us with a different one-loop effective potential 
\begin{eqnarray}
\tilde V_{\rm{eff}}^{(B)} = \frac{1}{64 \pi^2} m^4 \rho^4 \Bigl( \log \frac{m^2}{\mu^2} - \frac{1}{2}  \Bigr). 
\label{Diff-EP}  
\end{eqnarray}
It is worthwhile to stress that it is not the effective potential (\ref{Ren-effective potential}) but (\ref{Diff-EP})
that coincides with Eq. (\ref{Effective potential}), which is the standard and well-known result.

To understand the difference between the two effective potentials, let us notice that the actions (\ref{Scalar action}) 
and (\ref{Scalar action 2}) are invariant under a {\it{fake}} Weyl transformation 
\begin{eqnarray}
\bar g_{\mu\nu} \rightarrow \omega(x)^2 \bar g_{\mu\nu},  \qquad
\bar \varphi \rightarrow \omega(x)^{-1} \bar \varphi, \qquad
\rho \rightarrow \omega(x)^{-1} \rho, 
\label{Fake}  
\end{eqnarray}
since both $g_{\mu\nu}$ and $\varphi$ are trivially invariant under (\ref{Fake}). However, we easily find that
the former cutoff regularization breaks this fake symmetry while the latter cutoff regularization preserves it. 

Normally, without anomalies an effective action should possess the same contents of symmetries as its classical action
even if the symmetry is a fake symmetry as in (\ref{Fake}).\footnote{We think that the fake Weyl transformation (\ref{Fake})
is free from conformal anomaly as in the Weyl invariant scalar-tensor gravity, which was
extensively discussed in Ref. \cite{Tsamis}.}  In fact, we can verify that the effective action corresponding to the
effective potential (\ref{Diff-EP}) is invariant under the fake Weyl transformation (\ref{Fake}). This can be exhibitted 
by rewriting the conformal factor $\rho^4$ as $\sqrt{-g}$:
\begin{eqnarray}
\tilde \Gamma (g_{\mu\nu} ) \equiv -  \int d^4 x \tilde V_{\rm{eff}}^{(B)}
= - \frac{1}{64 \pi^2} m^4 \int d^4 x \sqrt{-g} \Bigl( \log \frac{m^2}{\mu^2} - \frac{1}{2}  \Bigr),
\label{EA-class1}  
\end{eqnarray}
where $\sqrt{-g} = \rho^4$ was used. This effective action is certainly invariant under the fake Weyl 
transformation (\ref{Fake}) as well as the general coordinate transformation since it is expressed in terms of the metric 
$g_{\mu\nu}$, which is trivially invariant under the fake Weyl transformation (\ref{Fake}). 

On the other hand, an effective action corresponding to the effective potential (\ref{Ren-effective potential}) is 
not invariant under the fake Weyl transformation (\ref{Fake}). Of course, we can rewrite the effective action
as an expression which is invariant under the fake Weyl transformation (\ref{Fake}), but then
the general coordinate symmetry is violated since
\begin{eqnarray}
\Gamma (g_{\mu\nu} ) \equiv -  \int d^4 x V_{\rm{eff}}^{(B)}
= - \frac{1}{64 \pi^2} m^4 \int d^4 x \sqrt{-g} \Bigl( \log \frac{m^2 (-g)^{\frac{1}{4}}}{\mu^2} - \frac{1}{2}  \Bigr),
\label{EA-class2}  
\end{eqnarray}
where the presence of the factor $(-g)^{\frac{1}{4}}$ breaks the general coordinate symmetry.
 
In this sense, we can regard not $V_{\rm{eff}}^{(B)}$ but $\tilde V_{\rm{eff}}^{(B)}$ as a physically plausible
effective potential for the classical theory defined by the action (\ref{Scalar action}). To put it differently, 
the effective potential $V_{\rm{eff}}^{(B)}$ defines a quantum theory which is different from the classical 
theory (\ref{Scalar action}) in the sense that $\rho$ does not have the meaning of the conformal 
factor any longer but a mere scalar field, which might be called ${\it dilaton}$ since it is created from 
a symmetry breaking of the fake Weyl symmetry.  This observation will be verified from a proof 
mentioned in Section 4.

\section{Fermionic spinor fields}

Before doing so, for the sake of completeness and recent works \cite{Oda-Kin1, Oda-Kin2}, let us consider 
the case of a massive Dirac spinor in a curved background. We will find the similar problem, i.e., the existence 
of two different effective potentials depending on the regularization procedure. The action of the massive 
Dirac spinor is given in a curved background: 
\begin{eqnarray}
S_F (\Psi, e_\mu^a ) &=& \int d^4 x \, e ( i \bar \Psi e_a^\mu \gamma^a D_\mu \Psi - m \bar \Psi \Psi ),
\label{Spinor action}  
\end{eqnarray}
where the covariant derivative is defined as $D_\mu \Psi = ( \partial_\mu + \omega_\mu ) \Psi$
with the spin connection $\omega_\mu^{ab}$. 

As in Eq. (\ref{Conformal factor}), when we define a conformal factor $\rho(x)$ by 
\begin{eqnarray}
e_\mu^a = \rho(x) \bar e_\mu^a,   \qquad
\Psi = \rho(x)^{-\frac{3}{2}} \psi,
\label{Conformal factor 2}  
\end{eqnarray}
the action (\ref{Spinor action}) can be written as
\begin{eqnarray}
S_F^\prime (\psi, \bar e_\mu^a, \rho ) = \int d^4 x \, \bar e [ \bar \psi ( i \slashed{D} - m \rho ) \psi + \dots ],
\label{Spinor action 2}  
\end{eqnarray}
where the ellipses again denote terms including derivatives of $\rho$.

Following a similar argument to the case of the scalar field, we find that an one-loop effective action is given by
\begin{eqnarray}
\Gamma_F ( \rho ) &=& - i {\rm{tr}} \log ( i \slashed{D} - m \rho )
\nonumber\\
&=& - i \int d^4 x \langle x | {\rm{tr}} \log ( i \slashed{D} - m \rho ) | x \rangle 
\nonumber\\
&=& - i \int d^4 x \int \frac{d^4 p}{(2 \pi)^4} {\rm{tr}} \log ( \slashed{p} - m \rho )
\nonumber\\
&\equiv& - \int d^4 x \, V_{\rm{eff}}^{(F)},
\label{Effective action F}  
\end{eqnarray}
where we have set $\omega_\mu = 0$ in the third equality.

This effective potential can be computed as follows:
\begin{eqnarray}
V_{\rm{eff}}^{(F)} &=& i \int \frac{d^4 p}{(2 \pi)^4} {\rm{tr}} \log ( \slashed{p} - m \rho )
\nonumber\\
&=& i \int \frac{d^4 p}{(2 \pi)^4} {\rm{tr}} \log \gamma^5 ( \slashed{p} - m \rho ) \gamma^5
\nonumber\\
&=& i \int \frac{d^4 p}{(2 \pi)^4} {\rm{tr}} \log ( - \slashed{p} - m \rho ) 
\nonumber\\
&=& \frac{1}{2} i \int \frac{d^4 p}{(2 \pi)^4} {\rm{tr}} [ \log ( \slashed{p} - m \rho ) + \log ( - \slashed{p} - m \rho ) ]
\nonumber\\
&=& 2 i \int \frac{d^4 p}{(2 \pi)^4} \log ( p^2 + m^2 \rho^2 ), 
\label{Effective potential F}  
\end{eqnarray}
where we have used the Clifford algebra, $\{ \gamma^a, \gamma^b \}  = - 2 \eta^{ab}$.
This effective potential has the same form as that of the scalar field, so we can arrive at
the renormalized one-loop effective potential for the conformal factor
\begin{eqnarray}
V_{\rm{eff}}^{(F)} = - \frac{4}{64 \pi^2} m^4 \rho^4 \Bigl( \log \frac{m^2 \rho^2}{\mu^2} - \frac{1}{2}  \Bigr).
\label{Ren-effective potential 2}  
\end{eqnarray}
In the above derivation, we have used the cutoff regularization given by
$\bar g^{\mu\nu} p_{E\mu} p_{E\nu} = \delta^{\mu\nu} p_{E\mu} p_{E\nu} < \Lambda^2$.

As in the scalar field, if we use a different momentum cutoff by using the metric 
$g^{\mu\nu}$, i.e., $g^{\mu\nu} p_{E\mu} p_{E\nu} = \rho^{-2} \delta^{\mu\nu} p_{E\mu} p_{E\nu} < \Lambda^2$,
we can obtain the one-loop effective potential
\begin{eqnarray}
\tilde V_{\rm{eff}}^{(F)} = - \frac{4}{64 \pi^2} m^4 \rho^4 \Bigl( \log \frac{m^2}{\mu^2} - \frac{1}{2}  \Bigr).
\label{Diff-EP 2}  
\end{eqnarray}
In this case as well, the physically plausible effective potential is given by $\tilde V_{\rm{eff}}^{(F)}$,
but not $V_{\rm{eff}}^{(F)}$.

As a final comment, it is straightforward to calculate the effective potential for the conformal factor where
matter fields are taken to be the abelian electromagnetic field or the nonabelian Yang-Mills field. However,
these fields are invariant under the Weyl transformation, or equivalently a local scale transformation, so we have no
effective potential for the conformal factor. This situation remains unchanged even if there is a conformal
anomaly since the conformal anomaly is expressed in terms of the Riemannian tensors.

\section{Effective potential from $GL(4)$ symmetry}

In the previous two sections, we have reached somewhat a strange conclusion: Although we have started with
two theories which are classically equivalent to each other, we have two different quantum theories irrespective of 
any anomalies. The source of this strange conclusion is obvious: The metric field plays a dual role, one of which is
a dynamical variable and the other is a geometrical object. The geometrical role enters in a theory in defining
the momentum squared. Actually, with the presence of a real scalar field $\phi$, there is an ambiguity in the choice 
of the metric, that is, the original metric $g_{\mu\nu}$ or a generalized metric
where the metric $g_{\mu\nu}$ is multiplied by some function of a scalar field, that is, $f (\phi) g_{\mu\nu}$. 
In the previous examples, the choice of the original metric $g_{\mu\nu}$ produces an effective potential 
which is quartic in the conformal factor whereas that of the modified metric $\bar g_{\mu\nu} = \rho^{-2} g_{\mu\nu}$ 
leads to a Coleman-Weinberg like effective potential. 
     
In Section 2, we have insisted that the physically plausible effective potential must take the form $V_{\rm{eff}}
= C \rho^4$ with a certain constant $C$ if we would like to regard $\rho$ as the conformal factor of the metric. 
To support our opinion, in this section, we will prove that the effective potential 
for conformal factor of the metric in the presence of a scalar field $\varphi$ necessarily takes the form of 
$\sqrt{-g} V (\varphi)$ on the basis of global $GL(4)$ symmetry.\footnote{T. Kugo has already presented 
an alternative proof which utilizes the change of momentum variables from a curved space-time to a flat Minkowski 
space-time in an effective potential \cite{Kugo}.}  

For this aim, we assume that the vacuum is translationally invariant, by which all fields are constants in space-time.
For such constant fields, the general coordinate transformation is reduced to a global $GL(4)$ 
transformation
\begin{eqnarray}
x^\mu \rightarrow  x^{\prime\mu} = ( M^{-1} )^\mu\,_\nu x^\nu,
\label{GL4}  
\end{eqnarray}
where $M^\mu\,_\nu$ is a constant $4 \times 4$ matrix obeying $\det M \neq 0$. 
Under the $GL(4)$ transformation, the constant fields and the Lagrangian density transform as
\begin{eqnarray}
&{}& g_{\mu\nu} \rightarrow  g_{\mu\nu}^\prime = g_{\alpha\beta} M^\alpha\,_\mu M^\beta\,_\nu, \qquad
\varphi^\prime \rightarrow \varphi,
\nonumber\\
&{}& {\cal{L}} (g, \varphi)  \rightarrow {\cal{L}}^\prime (g^\prime, \varphi^\prime) = \det M \cdot {\cal{L}} (g, \varphi).
\label{GL4-field}  
\end{eqnarray}
Here note that the scalar field $\varphi$ is invariant under the $GL(4)$ transformation. 

With the infinitesimal $GL(4)$ transformation, $M^\mu\,_\nu = \delta^\mu_\nu + \delta M^\mu\,_\nu$
with $| \delta M^\mu\,_\nu | \ll 1$, the constant fields and the Lagrangian density transform as
\begin{eqnarray}
\delta g_{\mu\nu} = \delta M_{\mu\nu} + \delta M_{\nu\mu}, \qquad
\delta \varphi = 0, \qquad
\delta {\cal{L}} = {\rm{tr}} \delta M \cdot {\cal{L}},
\label{GL4-field 2}  
\end{eqnarray}
where ${\rm{tr}} \delta M \equiv g^{\mu\nu} \delta M_{\mu\nu}$. Note that the last transformation implies that
the Lagrangian density indeed transforms as a density under the $GL(4)$ transformation.

Given the constant fields, under the infinitesimal $GL(4)$ transformation, the Lagrangian density transform as
\begin{eqnarray}
\delta {\cal{L}} = {\rm{tr}} \delta M \cdot {\cal{L}} = \frac{\partial {\cal{L}}}{\partial g_{\mu\nu}} 
( \delta M_{\mu\nu} + \delta M_{\nu\mu} ), 
\label{GL4-Lag}  
\end{eqnarray}
where we have used $\delta \varphi = 0$ in Eq. (\ref{GL4-field 2}). Then, this equation is simply solved to be
\begin{eqnarray}
{\cal{L}} = - \sqrt{-g} V (\varphi), 
\label{Sol}  
\end{eqnarray}
where $V (\varphi)$ is a function depending on only the scalar field $\varphi$. This proof
does not rely on perturbation theory so the result holds even in the non-perturbative regime.\footnote{A
similar argument based on a global $GL(4)$ symmetry was used in proving what we call the Weinberg no-go
theorem of the cosmological constant \cite{Weinberg, Oda1, Oda2, Oda3}.}

Now let us apply this result to the present situation where the scalar field $\varphi$ corresponds to 
the conformal factor $\rho$. Let us recall that in the effective potential $\rho$ is treated as a constant.
With $g_{\mu\nu} = \rho^4 \eta_{\mu\nu}$, Eq. (\ref{Sol}) leads to
\begin{eqnarray}
V_{\rm{eff}} (\rho, \varphi) = \rho^4 V (\varphi), 
\label{Sol-Eff}  
\end{eqnarray}
where we have defined the ``effective potential'', $V_{\rm{eff}} (\rho, \varphi) = - {\cal{L}}$.
If $V (\varphi)$ is positive definite, $\rho = 0$ is only the minimum of the effective potential,
thereby meaning that the metric is degenerate at low energies and the inverse metric does not 
exist, i.e., $\langle g_{\mu\nu} \rangle = 0$. It is worthwhile to notice that the effective
potential never takes the form like 
\begin{eqnarray}
V_{\rm{eff}} (\rho, \varphi) = \rho^4 ( a \log \rho + b ), 
\label{Sol-Eff 2}  
\end{eqnarray}
where $a, b$ are constants depending on the constant $\rho$.
 
Then, a natural question arises whether quantum effects of matter and gravitational fields do not always 
drive the vacuum expectation value of the metric to be non-degenerate.  To answer this question, let us
note that the key ingredient in our derivation of Eq. (\ref{Sol}) is the presence of a global $GL(4)$ symmetry 
for constant fields. It is known that in a theory of quantum gravity the $GL(4)$ symmetry is spontaneously broken 
to the Lorentz symmetry, and consequently the exactly massless graviton emerges as a Nambu-Goldstone particle 
associated with this spontaneous symmetry breakdown \cite{NO}. In fact, if $M^\mu\,_\nu$ belongs to 
a generator of the Lorentz group $SO(1,3)$, we have 
\begin{eqnarray}
\delta g_{\mu\nu} = \delta M_{\mu\nu} + \delta M_{\nu\mu} = 0, 
\label{Lorentz}  
\end{eqnarray}
since $\delta M_{\mu\nu}$ is an antisymmetric matrix in case of the Lorentz group.
In such a broken phase of the $GL(4)$ symmetry, our derivation based on the $GL(4)$ symmetry
does not make sense so there could be a possibility that a nontrivial effective potential 
appears, which drives the vacuum expectation value of the metric or vierbein to be non-degenerate.

\section{Conclusion}

In this article, we have studied the issue of the effective potential of the conformal factor of the metric 
generated by quantized matter fields. It has been known that there are two kinds of the effective potentials,
one of which produces a non-vanishing vacuum expectation value (VEV) of the conformal factor
and the other does not exhibit such a feature, that is, a vanishing VEV. If the former effective potential
were true, we could show that we have a non-degenerate metric at low energies even if there is 
a denegerate metric, which is required for the change of the space-time topology, at high energies. 

Based on a gobal $GL(4)$ symmetry, which exists in the formulation of effective potentials, we have
proved that the effective potential for the conformal factor always takes the form of
$V_{\rm{eff}} (\rho, \varphi) = \rho^4 V (\varphi)$, but not $V_{\rm{eff}} (\rho, \varphi) = \rho^4 
( a \log \rho + b )$. This proof clearly implies that we cannot show the existence of the non-degenerate
metric at low energies from quantum effects associated with matter fields.  Moreover, our proof
suggests that we might be able to show the existence of the non-degenerate metric in
a quantum gravity where the global $GL(4)$ symmetry is spontaneously broken to the Lorentz
symmetry.

\begin{flushleft}
{\bf Acknowledgements}
\end{flushleft}

We are grateful to T. Kugo for valuable discussions.
This work is supported in part by the JSPS Kakenhi Grant No. 21K03539.


\end{document}